\documentstyle[12pt]{article}
\newcommand{\be}{\begin{equation}}
\newcommand{\ee}{\end{equation}}
\newcommand{\bea}{\begin{eqnarray}}
\newcommand{\eea}{\end{eqnarray}}
\newcommand{\inn}{\!\cdot\!}
\newcommand{\norm}[1]{\raise.3ex\hbox{:}#1\raise.3ex\hbox{:}}

\begin{document}
\pagestyle{plain}
\setcounter{page}{1}
\newcounter{bean}
\baselineskip16pt


\begin{titlepage}
\begin{flushright}
PUPT-1744\\
ITEP/TH-41/97\\
hep-th/9711112
\end{flushright}

\vspace{7 mm}

\begin{center}
{\huge Dynamics of $(n, 1)$ Strings }

\end{center}
\vspace{10 mm}
\begin{center}
{\large  
S.~Gukov\footnote{On leave from the Institute of
Theoretical and Experimental Physics and L.D.~Landau
Institute for Theoretical Physics}, I.R.~Klebanov and
A.M.~Polyakov\\
}
\vspace{3mm}
Joseph Henry Laboratories\\
Princeton University\\
Princeton, New Jersey 08544
\end{center}
\vspace{7mm}
\begin{center}
{\large Abstract}
\end{center}
\noindent
An $(n, 1)$ string is a bound state of a D-string and $n$
fundamental strings. It may be described by a D-string with
a world volume electric field turned on. As the electric field 
approaches its critical value, $n$ becomes large.
We calculate the 4-point function for transverse oscillations of
an $(n, 1)$ string, and the two-point function for massless closed
strings scattering off an $(n, 1)$ string. In both cases we find
a set of poles that becomes dense in the large $n$ limit.
The effective tension that governs the spacing of these poles is the
fundamental string tension divided by $1+(n\lambda)^2$, where
$\lambda$ is the closed string coupling. We associate this effective 
tension with the open strings attached to the $(n, 1)$ string, 
thereby governing its dynamics. We also argue that the effective 
coupling strenth of these open strings is reduced by the electric 
field and approaches zero in the large $n$ limit.
\vspace{7mm}
\begin{flushleft}
November 1997

\end{flushleft}
\end{titlepage}


\renewcommand{\epsilon}{\varepsilon}
\def\fixit#1{}
\def\comment#1{}
\def\equno#1{(\ref{#1})}
\def\equnos#1{(#1)}
\def\sectno#1{section~\ref{#1}}
\def\figno#1{Fig.~(\ref{#1})}
\def\D#1#2{{\partial #1 \over \partial #2}}
\def\df#1#2{{\displaystyle{#1 \over #2}}}
\def\tf#1#2{{\textstyle{#1 \over #2}}}
\def\Leff{L_{\rm eff}}
\def\at{{\tilde \alpha}}
\def\a{{\alpha}}
\def\pt{{\tilde \psi}}
\def\p{{\psi}}
\def\pa{\partial}
\def\lm{\lambda}
\def\cd{\cal D}
\def\ce{\cal E}
\def\ep{\varepsilon}
\def\cf{{\cal F}}
\def\nn{\nonumber}
\def\cm{{\cal M}}


\section{Introduction}

Type IIB string theory is believed to possess a non-perturbative
SL(2,Z) symmetry \cite{HT,Wit}. This implies that the theory contains
various types of strings, labeled by relatively prime integers $p$
and $q$. In this scheme the fundamental string is $(1, 0)$ while
the basic RR-charged (Dirichlet) string is $(0,1)$. The classical
solutions of type IIB supergravity, which correspond to
the entire SL(2,Z) multiplet of $(p,q)$ strings were constructed
in \cite{S}. In the D-brane description \cite{dlp,polch}, 
the $(0,1)$
string is the basic D-string characterized by the Neumann boundary
conditions for directions $0$ and $1$, and by Dirichlet boundary
conditions along the remaining directions. In a subsequent development,
Witten identified a $(p,q)$ string with a bound state of $p$ fundamental
and $q$ D-strings \cite{W}. This bound state is a particular state in the
$U(q)$ gauge theory which describes $q$ parallel D-strings. 

In \cite{Li,CK,ASJ} the $(p,q)$ bound states were studied from the point
of view of the non-linear Born-Infeld action \cite{BI},
which describes the open strings that define the D-brane dynamics.
It was found that the correct BPS formula for the $(p,q)$ string
tension,
\be
T_{p,q} = T_{1,0}\sqrt {p^2 + {q^2\over \lambda^2}}
\ ,\ee
follows from a straightforward quantization of the non-linear
action for the collective coordinate describing 
the electric field \cite{CK}.
The situation is particularly simple for $q=1$
where the theory is abelian, and the action is known
in detail. Here the $(n,1)$ string corresponds to
a state with $n$ units of electric flux.
If we expand the exact mass formula for weak coupling $\lambda$,
we find
\be
T_{n,1}= T_{0,1} + {1\over 2}\lambda n^2 T_{1, 0} + O(\lambda^3)
\ .\ee
This means that for small $\lambda$ the $n$ fundamental strings lose
their energy almost entirely in the process of binding to a single
D-string. Note, however, that the extra energy due to the $n$
bound fundamental string grows as $n^2$. 
It may be interesting,
therefore, to study what happens for large values of $n$ (this is where
the electric field approaches its critical value \cite{CK}).
This situation
was first considered by H. Verlinde \cite{V}. If the large $n$ limit is
taken first then the mass formula may be expanded as
\be \label{expand}
T_{n,1}= n T_{1,0} + {T_{0, 1}\over 2 n\lambda} + \ldots
\ee
This implies that the effective tension of a D-string bound to a large
number $n$ of fundamental strings has decreased by a factor 
$2n\lambda$ \cite{V}. Thus, it is possible that the low-lying
excitations of the bound state are described by a string with
rescaled tension.

One of the motivations in \cite{V} was that the S-duality maps a $(n,1)$
string into a $(1,n)$ string. Thus, by studying a single D-string in
the near-critical electric field we may learn something about
the large $n$ limit of $N=8$ supersymmetric $U(n)$ gauge theory in
$1+1$ dimensions. Another reason to be interested in this problem
is the theory of confining strings describing gauge theories.
It was shown in \cite{AP} that in this case one expects tensionless
open strings interacting with tensile closed strings.
In this paper we show that precisely this situation occurs for a D-string
in the near-critical world volume electric field.
This fact has a simple qualitative explanation. Indeed,
the open strings have charges attached to their ends.
The electrostatic energy in a constant electric field is 
proportional to the length of the string. When the field is just below
critical, it reduces the effective string tension almost
to zero.\footnote{A related effect in the type I theory
is that, in the critical
electric field, the tunneling barrier for pair creation of 
open strings disappears \cite{BP}.} 
The field above critical would tear the open string apart.

In this paper we probe the large $n$ limit of the $(n,1)$ bound state
with dynamical calculations which go beyond the BPS limit.
We calculate the 4-point function for transverse oscillations of the
$(n,1)$ string and find that the poles indeed become dense in the
limit where $n\lambda\rightarrow \infty$. The relevant tension
which governs the spacing of the poles is 
\be \label{rescaled}
T_{eff} = {T_{1, 0}\over  1+(n\lambda)^2} 
\ .\ee
In the large $n$ limit this is equal to the fundamental string tension
reduced by the factor $(n\lambda)^2$.
The strings that lose their tension here are the open strings
attached to the $(n,1)$ string, i.e. the objects that
define their dynamics. Note that (\ref{rescaled}) is different
from the rescaling factor in the effective tension of the D-string,
\be \label{Drescale}
T_{Deff} \sim {T_{1,0}\over   n\lambda^2}\ , \ee
which appears in the BPS formula.
We believe that the apparent difference is due to the fact that
the string coupling is rescaled as well:
$$ \lambda_{eff} = \lambda {1\over \sqrt {1+ (n\lambda)^2}}
\ ,$$
so that
$$ T_{Deff} = {T_{eff}\over \lambda_{eff}}
$$
scales according to (\ref{Drescale}).
This rescaling affects only the interaction strength of the open
strings, which move in $1+1$ dimensions along the bound state.
The interaction strength of the closed strings, which move in the bulk,
is independent of the electric field.

We conclude the paper by indicating how to
probe the $(n,1)$ string with massless closed strings
incident from the outside. The rescaled open string tension
(\ref{rescaled}) can be seen in these amplitudes as well. 
The amplitudes also show that, as expected,
the tension of the closed strings
propagating in the bulk is unaffected by the electric field 
on the D-string.

\section{Setup}

We study a bound state of $n$ fundamental strings with a D(irichlet)
string \cite{W} in type IIB string theory. From the conformal field
theory point of view we need to
introduce the following boundary term into the action \cite{W,CK,ASJ},
\be \label{bcft}
S_b = \oint d\sigma 
E X^0 {\pa \over \pa \sigma} X^1
\ ,
\ee
where $X^1$ is the compact direction of length
$l$ over which the string is wrapped.
$E= \dot A_1$, and the lagrangian for the collective coordinate
$A_1$ is given by the DBI action,
\be
 L= -l {T_{1,0} \over \lambda} \sqrt {1-E^2}\ .
\ee
On compact $X^1$ the
momentum conjugate to $A_1$ is quantized. Taking $n$ quanta of
it we get the $(n,1)$ string tension \cite{CK}
\be
T_{n,1} = T_{1,0} \sqrt{ n^2 + {1 \over \lm^2}}
\ .\label{tens}
\ee
For reasons explained in the introduction, we are mostly
interested in the large $n$ limit.
The expression for the electric field is \cite{CK} 
\bea
E={n \lm \over \sqrt{1+ n^2 \lm^2}}
\ .\nonumber
\eea
Thus, in the
large $|n|$ limit $E$ tends to its critical value
$E_{c}= \pm 1$.

The boundary interaction (\ref{bcft}) assigns a specific
set of boundary conditions on the real axis. It turns out
that the boundary conditions have a linear form,
\be
\label{bc}
\tilde X^{\mu}= {D^{\mu}}_{\nu}X^{\nu}\ ,
\ee
where $\tilde X^{\mu}$ and $X^\mu$ are the antiholomorphic
and the holomorphic parts of the field respectively. While it is possible
to work on the upper half-plane ${\cal H}^{+}$, 
it is more convenient to use the doubling 
trick \cite{GM,HK} where the holomorphic part of the field,
$X^\mu(z)$, is extended to the entire complex plane 
in the following way:
\be 
\left\{ \begin{array}{cc}
X^{\mu} (z) \rightarrow X^{\mu}(z)  & z \in {\cal H}^{+}\ , \\
\tilde X^{\mu}(z) \rightarrow {D^{\mu}}_{\nu}X^{\nu} (z)
& z \in {\cal H}^{-} \ .
\end{array} \right.
\label{replace}
\ee
This replacement
allows us to express all correlators in terms of
holomorphic variables only.
All we need is to determine the form of the matrix ${D^{\mu}}_{\nu}$.
In fact, its form is well known from the analysis of the boundary
state.

The bosonic part of the boundary state (the fermionic part has analogous
form) is \cite{CK,DV1,DV2}:
\be
|B \rangle = T_{1,0} \sqrt{1-E^2} \prod_{j=2}^9 \delta(X^j)
\exp \left[ - \sum_{n=1}^{\infty} {1\over n}\a^{\mu}_{-n} D_{\mu \nu}(E)
\at^{\nu}_{-n} \right] |0\rangle
\label{b}
\ee
where the matrix ${D^{\mu}}_{\nu}$ has the form of a Lorentz boost,
\bea
{D^{\mu}}_{\nu} = \pmatrix { {\cd} & 0\cr 0 & -1\cr}
& \rm{where} &
{{\cd}^{\mu}}_{\nu} = {1\over 1-E^2} \pmatrix { 1+E^2 & 2E\cr 2E & 1+E^2\cr}
\ .\label{d}
\eea

It is quite clear that
the matrix ${D^{\mu}}_{\nu}$ entering the boundary state is the same 
matrix as the one in (\ref{bc}) and
(\ref{replace}). 
This is because the boundary state satisfies
\be \tilde \alpha_n^\mu 
|B\rangle = {D^{\mu}}_{\nu} \alpha_{-n}^\nu
|B\rangle
\ ,\ee
which implies the boundary condition (\ref{bc}).

Though throughout the paper we use conformal field theory
techniques and mainly apply the doubling trick, all the
results can be derived also from the boundary state (\ref{b}).
Let us review some calculations where the
{\it tensionless strings} can be excited.


\section{The four-point amplitude}

The simplest nontrivial example where one can probe the tensionless
strings is the four-point amplitude for NS open strings \cite{HK}:
\begin{eqnarray}
A_{4}(\xi_1,p_1;\xi_2,p_2;\xi_3,p_3;\xi_4,p_4) \sim {1\over \lambda}
\sqrt{1-E^2} \int \{ d\sigma_1 d\sigma_2 d\sigma_3 d\sigma_4 \} \nonumber \\
\langle \xi_1\inn V_{0}(p_1,\sigma_1)\,
\xi_2\inn V_{0}(p_2,\sigma_2) \, 
\xi_3\inn V_{-1}(p_3,\sigma_3) \,
\xi_4\inn V_{-1}(p_4,\sigma_4) \rangle
\nonumber
\end{eqnarray}
The vertex operators for scalar particles (the transverse modes of the
string) have the form:
\begin{eqnarray}
V_{-1}^j(p_1,z) = e^{-\phi(z)}\,\psi^j (z)\,e^{ip_1\cdot
X(z)}
\label{vert} \\
V_0^j (p_2,z) =\left(\partial X^j (z)+ip_2\inn \psi(z)\psi^j (z)
\right)\,e^{ip_2\cdot X(z)} \nonumber 
\end{eqnarray}
where $j=2, \ldots, 9$, while the momenta $p^\alpha$ are longitudinal,
$\alpha=0,1$.
$z$ must lie on the real axis. For the $(0,1)$ string, the
holomorphic and antiholomorphic parts of $X$ are identical; hence,
$X(z, {\bar z})$ may be replaced by twice the holomorphic part:
\bea
X^{\alpha}(z,{\bar z}) \rightarrow 2 X^{\alpha} (z)
\nonumber
\ .\eea
Thus, all we need to do is replace
\bea
p^{\alpha} \rightarrow 2 p^{\alpha} 
\nonumber
\eea
in the usual type I 4-point amplitude.
In the $(n,1)$ case we must be more careful with $X^0$ and $X^1$. Because
of the boost, the appropriate replacement is:
\be
p^{\alpha} \rightarrow p^{\alpha} + {{\cd}^{\alpha}}_{\beta} p^{\beta}
\label{repl1}
\ .\ee
It turns out that
the amplitude includes a phase depending on the
ordering of the vertex operators, which originates from the second
term in the Green function on the boundary:
\be\label{Green}
\langle X^\alpha (\sigma_1) X^\beta (\sigma_2) \rangle =
-{1\over 1-E^2}\eta^{\alpha\beta} \ln |\sigma_1-\sigma_2|
+ i{\pi\over 2}
 {E\over 1-E^2}\epsilon^{\alpha\beta} {\rm sgn} (\sigma_1-\sigma_2)
\ .\ee
Such phases are not present for $E=0$. Thus, they constitute a
new feature of the $(n,1)$ dynamics.

Let us consider the 4-point function for back-scattering.
In the center of mass frame the momenta are
\bea
p^{\alpha}_1 = \pmatrix { p \cr p\cr}\ , \qquad
p^{\alpha}_2 = \pmatrix { p \cr -p\cr}\ , \qquad
p^{\alpha}_3 = \pmatrix { -p \cr -p\cr}\ , \qquad
p^{\alpha}_4 = \pmatrix { -p \cr p\cr}\ .
\nonumber
\eea
Let us position $V_2$ and $V_4$ at 0 and 1, $V_3$ at $\infty$, and
integrate over the position of $V_1$ from 0 to 1.
The integral is of the well-known type I form:
\bea
 {1\over \lambda}
\sqrt{1-E^2} {\Gamma(p_1\inn p_2) \Gamma(p_1\inn p_4) 
\over \Gamma(p_1\inn p_2+ p_1\inn p_4+1)}
K(\xi_1,p_1;\xi_2,p_2;\xi_3,p_3;\xi_4,p_4)
\nonumber
\eea
where $K$ is the kinematic factor:
\begin{eqnarray}
K(\xi_1,p_1;\xi_2,p_2;\xi_3,p_3;\xi_4,p_4) =
-p_2\inn p_3\,p_2\inn p_4\ \xi_1\inn\xi_2\,\xi_3\inn\xi_4
%
-{1 \over 4}p_1\inn p_2\,
(\xi_1\inn p_4\,\xi_3\inn p_2\,\xi_2\inn\xi_4 +
\nonumber\\
+\xi_2\inn p_3\,\xi_4\inn p_1 \,\xi_1\inn\xi_3
%
+ \xi_1\inn p_3\,\xi_4\inn  p_2\,\xi_2\inn\xi_3
 +\xi_2\inn p_4\,\xi_3\inn p_1\,\xi_1\inn\xi_4)+
\nonumber\\
%
+\Big\{1,2,3,4\rightarrow 1,3,2,4\Big\} 
+\Big\{1,2,3,4\rightarrow 1,4,3,2\Big\}
\nonumber
\end{eqnarray}
but with all momenta replaced according to (\ref{repl1}).
We should also take into account the phases from
the second term in (\ref{Green}).
For $V_2$ at 0 and $V_4$ at 1, the phase is 
$\exp\left (2\pi i {E p^2\over 1-E^2}\right )$.
For $V_4$ at 0 and $V_2$ at 1, the phase is
$\exp\left (-2\pi i {E p^2\over 1-E^2}\right )$.
The complete s-channel amplitude is
\bea
A_4 & \sim &- {\sqrt{1-E^2}\over \lambda}
\cos\left (2\pi {E p^2\over 1-E^2}\right )
s^2 \Gamma(s) \Gamma(-s)
\xi_1\cdot \xi_3 \xi_2 \cdot \xi_4 \nonumber \\ & = &
- {\sqrt{1-E^2}\over \lambda}
\cos\left (2\pi {E p^2\over 1-E^2}\right )
  {\pi s\over 
\sin(\pi s)} \xi_1\cdot \xi_3 \xi_2 \cdot \xi_4
\eea
with $s$ given by
\bea
s=p_{1 \mu} \left( {\delta^{\mu}}_{\lm} + {D^{\mu}}_{\lm} \right)
\left( {\delta^{\lm}}_{\nu} +  {D_{\nu}}^{\lm} \right) p_2^{\nu}=
-{8p^2 \over 1-E^2 } 
\label{kinvar}
\eea
Comparing to the $n=0$ case we see that $s$ is rescaled by the factor 
$${1\over 1-E^2}=  1 + (n\lambda)^2
\ .$$
Thus, as $n\lambda$ increases the poles in the actual kinematical variable,
$-8 p^2$, become denser.
This is equivalent to a rescaling of the tension of the fundamental string,
\be \label{newten}
T_{eff}= T_{1,0} (1-E^2) = {T_{1,0}\over
1 + (n\lambda)^2}
\ .\ee
In the large $n$
limit the poles become infinitely dense,
and the process is governed by a kind of tensionless string.
The strings that
lose their tension are the fundamental open strings attached to the
$(n,1)$ string, i.e. the objects that define the dynamics
in the D-brane theory. Thus, the massive states of such strings
give rise to a tower of low energy excitations of the $(n,1)$
bound state, with spacing of order 
\be
\delta E \sim {1\over n\lambda\sqrt{\alpha'}}
\ee
in the large $n$ limit.

Is this the only effect of the electric field? We believe that the
answer is no: the string coupling gets rescaled as well.
Indeed, it is tempting to identify the disk amplitude with no insertions,
$$ {T_{1,0}\over\lambda} \sqrt{1-E^2}
\ ,$$
with $T_{eff}/\lambda_{eff}$. Using (\ref{newten}), we find
\be \label{newcoup}
\lambda_{eff}= \lambda \sqrt{1-E^2} = {1\over
\sqrt {{1\over\lambda^2} + n^2}}
\ .\ee
Another argument in favor of this rescaling is that an insertion of
an extra hole into the world sheet carries a factor $\sqrt{1-E^2}$
from the normalization of the boundary state, 
in addition to the obvious factor $\lambda$. Thus, $\lambda_{eff}$
is the effective hole counting parameter (it is the effective 
coupling constant squared for the open strings that move in
$1+1$ dimensions).\footnote{
Using the Born-Infeld action as the low-energy effective action
for the open strings, we have checked explicitly that the
$l$ loop correction to the 4-point function is of order
$\lambda_{eff}^l$ times the tree level 4-point function.}
Each handle, on
the other hand, introduces a factor $\lambda^2$ independent of $E$.
Thus, the interaction strength of the closed strings, which move
in the bulk, is independent of the electric field.

Note that in the large $n$ limit $\lambda_{eff}=1/n$ independent
of $\lambda$. Let us consider performing a type IIB S-duality 
transformation to the $(1,n)$ bound state described by the
supersymmetric $U(n)$ gauge theory. In this theory we expect that
\be g_{YMeff}^2 \sim {1\over \lambda_{eff}} \sim n\ .
\ee
The effective gauge coupling should be the `t Hooft coupling,
\be g_{YMeff}^2 = g_{YM}^2 n\ .
\ee
Thus, consistency requires that $g_{YM}$ is independent of $n$.
This is indeed the behavior of the $U(n)$ gauge coupling \cite{V}.
These arguments appear to support the scaling of the effective open
string coupling that we have found for the $(n,1)$ bound state.
 
We may think of $T_{eff}/\lambda_{eff}$ as the effective D-string
tension. With this definition, we find that the rescaling factor is
$1/(n\lambda)$. This is in qualitative agreement with the result
of \cite{V} but differs by a factor of 2.\footnote{
In fact, the two definitions
do not necessarily have to agree: in \cite{V} the energy per unit length
of a D-string bound to $n$ fundamental strings was obtained from
the energy of the bound state. 
We, on the other hand, are examining the lagrangian
per unit length. The qualitative agreement is manifest, however.}


\section{Other amplitudes}

Another example is scattering of a massless closed
string off a $(n,1)$ string. For this $1\rightarrow 1$
amplitude we can track the $D$ factors and
substitute (\ref{d}) into the final expression \cite{GM,HK}:
\be
A_2 \sim {1\over \lambda}\sqrt{1-E^2}
\int \{ dz_{1,2}\} \ep_{1 \mu \lm} {D^{\lm}}_{\nu}
\ep_{2 \sigma \eta} {D^{\eta}}_{\kappa} \langle
V^{\mu}_{-1}(p_1, z_1)
V^{\nu}_{-1}(Dp_1, {\bar z_1})
V^{\sigma}_0 (p_2, z_2)
V^{\kappa}_0 (Dp_2, {\bar z_2})
\rangle
\ .\label{a2}
\ee 
We have already defined the vertex operators in (\ref{vert}).
Calculations lead to the answer \cite{GM}:
\bea
A_2 ={1\over \lambda}
\sqrt{1-E^2} {\Gamma(s) \Gamma(t) \over \Gamma(1+s+t)}
\left( s a_1 - t a_2 \right)
\label{amplitude}
\eea
where $a_1$, $a_2$ are polarization-dependent kinematic
factors:
\begin{eqnarray}
a_1 & =&
{\rm Tr}(\varepsilon_1\inn D)\,p_1\inn \varepsilon_2 \inn p_1 
-p_1\inn\varepsilon_2\inn D\inn\varepsilon_1\inn p_2 
- p_1\inn\varepsilon_2\inn\varepsilon_1^T \inn D\inn p_1
 -p_1\inn\varepsilon_2^T \inn \varepsilon_1 \inn D \inn p_1 - \nonumber \\
&& - p_1\inn\varepsilon_2\inn\varepsilon_1^T \inn p_2 +
\frac{s}{2}\,{\rm Tr}(\varepsilon_1\inn\varepsilon_2^T) 
+\Big\{1\longleftrightarrow 2\Big\}
\nonumber \\
a_2 & = &
{\rm Tr}(\varepsilon_1\inn D)\,
(p_1\inn\varepsilon_2\inn D\inn p_2 
+ p_2\inn D\inn\varepsilon_2\inn p_1  
 +p_2\inn D\inn\varepsilon_2\inn D\inn p_2)
+p_1\inn D\inn\varepsilon_1\inn D\inn\varepsilon_2\inn D\inn p_2
- \nonumber \\
&&-p_2\inn D\inn\varepsilon_2\inn\varepsilon_1^T\inn D\inn p_1 
+\frac{s}{2}\,{\rm Tr}(\varepsilon_1\inn D\inn \varepsilon_2\inn D)
-\frac{s}{2}\,{\rm Tr}(\varepsilon_1\inn\varepsilon_2^T) - \nonumber \\
&&-\frac{s+t}{2}{\rm Tr}(\varepsilon_1\inn D) {\rm Tr}(\varepsilon_2\inn D)
+\Big\{1\longleftrightarrow 2 \Big\}\ \ .
\nonumber
\end{eqnarray}
The kinematical invariants are $t= p_1\cdot p_2$ and 
\bea
s=p_{1 \mu} {D^{\mu}}_{\nu} p_1^{\nu}= {2\over 1-E^2} p_{||}^2
\ ,
\eea
where
\be
p_{||}^2 = (p_1^1)^2- (p_1^0)^2= (p_2^1)^2- (p_2^0)^2
\ .
\ee
The rescaling of the kinematical variable $s$ is the same as in
(\ref{kinvar}). Thus, we find that 
$$ s= {2 p_{||}^2 \over 1+(n\lambda)^2}
\ ,
$$
which leads to dense poles in the actual kinematical
variable, $2 p_{||}^2$, for
large $n\lambda$.
These poles correspond to excitations of the $(n,1)$ string by attaching
to it excited open strings \cite{KT,GHKM}. 
Thus, we find further evidence that such
open string have effective tension (\ref{rescaled}) and become tensionless in
the large $n$ limit. It is also clear that the kinematical variable
$t$ is not rescaled; hence, as expected, the electric field does
not affect the tension of the closed strings propagating in the bulk.

Let us consider the simplest case: scattering of gravitons polarized
transversely to the string \cite{KT}. Then (\ref{amplitude})
simplifies to
\bea
A_2 \sim \sqrt {n^2+  {1\over \lambda^2} }
 {\Gamma(1+s) \Gamma(t) \over \Gamma(1+s+t)}
p_{||}^2 \epsilon_1\cdot \epsilon_2\ .
\eea
In the gravitational lensing ($t\rightarrow 0$) limit, this becomes
\be \label{lensing}
{1\over t} \sqrt {n^2+ {1\over \lambda^2} }
p_{||}^2 \epsilon_1\cdot \epsilon_2\ .
\ee
Note that the amplitude is proportional to the 
$(n,1)$ string tension. 
Now we compare with supergravity,
where the metric around the string is given by
\bea
ds^2 = A^{- {3 \over 4}} (-dt^2 + (dx^1)^2) + A^{1 \over 4}
d{\bf x}\cdot d{\bf x}\ , \label{metr}\\
A({\bf x})  = 1 + {T_{n,1}\over 3x^6}
\ .\nn
\eea
{}From the methods of \cite{GHKM} it is clear that,
since the long-range tail of the metric perturbation is proportional
to the tension, so is the coefficient of $1/t$.
Thus, we find complete agreement between (\ref{lensing})
and the corresponding supergravity result.

\section{Conclusion}

In this paper we investigated the physics of $(n,1)$ bound states
which consist of $n$ fundamental strings bound to a D-string.
In the D-brane theory the dynamics of the bound state
is described by open strings whose charged end-points move 
in $1+1$ dimensions subject to an electric field along the D-string.
Due to the electric field
the tension of these open strings, as well as their
coupling strength, become effectively
reduced and approach zero in the large $n$ limit.

The transverse size of the bound state is of the order
\be
\sqrt{\alpha'_{eff}} \sim {\sqrt{\alpha'}\over \sqrt{1-E^2}}
\sim \sqrt{\alpha'} \sqrt{ 1+ (n\lambda)^2 }
\ .
\ee
This means that, for large $n$, the transverse size grows as
$n\lambda \sqrt{\alpha'}$.
The growth with $n$ is indeed suggestive of
having $n$ constituents. Thus, the bound state becomes very thick,
even compared to the string scale.
In studying the excitations of this thick string, we were led to
the conclusion that their effective coupling constant squared,
$\lambda_{eff}$, decreases as $1/n$.
It would be interesting to understand this phenomenon better.

What happens in the large $n$ limit of
$(n,1)$ bound states can perhaps be regarded as a prototype for
the essential phenomenon in the theory of confining strings:
the open strings become tensionless while the
closed strings remain tensile \cite{AP}. We hope that this analogy
can be pursued further.

\section*{Acknowledgments}

We are grateful to H. Verlinde for useful discussions. 
This work was supported in part by the NSF grant
PHY-9600258, DOE grant DE-FG02-91ER40671,
the NSF Presidential Young Investigator Award PHY-9157482, and the
James S.{} McDonnell Foundation grant No.{} 91-48.



\end{document}